\documentclass[aps,pra,superscriptaddress,amsmath,amssymb,preprintnumbers,floatfix,showpacs,11pt]{revtex4}
\usepackage{amssymb}
\usepackage{epsfig}

\begin{document}
\title{ Single-atom entropy squeezing for two two-level atoms interacting
with a single-mode radiation field}
\author{ Faisal A. A. El-Orany }
\email{el_orany@yahoo.com}
 \affiliation{Department of Mathematics  and computer Science,
Faculty of Science, Suez Canal University 41522,
 Ismailia, Egypt}

\author{Wahiddin M. R. B.  }
 \affiliation{ Cyberspace Security
Laboratory, MIMOS Berhad, Technology Park Malaysia, 57000 Kuala
Lumpur, Malaysia}

\author{ Obada A-S  F }
 \affiliation{
 Mathematics Department, Faculty of Science, Al-Azhar
University, Nasr City, Cairo,  Egypt}

\date{\today}

\begin{abstract}
In this paper we consider a system of two  two-level atoms
interacting with a single-mode quantized electromagnetic field in
a lossless resonant cavity via $l$-photon-transition mechanism.
The field and the atoms are initially prepared in the coherent
state and the excited atomic states, respectively. For this system
we investigate the entropy squeezing, the atomic variances, the
von Neumann entropy and the atomic inversions for the single-atom
case. We show that the more the number of the parties in the
system the less the amounts of the nonclassical effects exhibited
in the entropy squeezing.
 The entropy squeezing can
give information on the corresponding von Neumann entropy. Also
 the nonclassical effects obtained form the asymmetric atoms
are greater than those obtained form the symmetric atoms. Finally,
the entropy squeezing  gives better information than the atomic
variances only for the asymmetric atoms.

\end{abstract}

  \maketitle

\section{Introduction}

The squeezed state of light is distinguished by a long-axis
variance of noise ellipse for one of its quadratures in the phase
space. This property has been used in many optical devices as well
as in the quantum information, e.g.  a power recycled
interferometer \cite{[1]}, a phase-modulated signal recycled
interferometer \cite{[2]}, quantum teleportation \cite{[3]},
cryptography \cite{[4],[5]} and dense coding \cite{[6]}. It is
worth pointing out that experiments for the quantum teleportation
have been successfully performed by means of the two-mode squeezed
vacuum states \cite{[7]}. Various  methods have been proposed for
the generation of squeezed states of the electromagnetic field and
some of them have been implemented, e.g., \cite{[8],[9]}.

The concept of squeezed states has been extended to atoms
\cite{wodk} and defined in a sense similar to that of the
radiation field.  In this respect, the atomic squeezing has been
obtained a great interest \cite{Aga} owing  to its potential
application in the high-resolution spectroscopy \cite{kita}, the
high-precision atomic fountain clock \cite{wine}, the
high-precision spin polarization measurements \cite{sor}, etc..
  Spin squeezing is another measure, which has been
applied to collected atoms \cite{sor}. Additionally, this measure
 has been used to quantify the entanglement in the
multi-atom systems \cite{cir}. Furthermore, spin squeezing has
been experimentally realized for an ensemble of V -type atoms
\cite{kuz}.
 In all these cases the atomic squeezing has been treated  in the
framework of  the Heisenberg uncertainty relations (HUR).
Nevertheless, the HUR cannot give sufficient information on the
atomic squeezing, in particular, when  the atomic inversion is
zero \cite{fang}. This difficulty  has been overcome by using the
entropic uncertainty relation (EUR) \cite{sanc}. In this regard
one has to use the  concept of  atomic entropy squeezing. More
details about this issue will be given in section 2. So far the
entropy squeezing technique has been applied to  the single
two-level atom interacting with a single mode or two modes, i.e.
the Jaynes-Cummings model (JCM) \cite{fang,{obo}}. In this paper
we apply this technique to  the system of two atoms
      interacting with the one-mode  electromagnetic field being in the
      $l$-th photon resonance
       with the atomic transition (TJCM).
The atoms and the field are initially prepared in the excited
atomic states and the coherent state, respectively. Moreover,  we
do not consider
  dissipation in the system, which generally leads to the degradation of the nonclassical
  effects. This means that  the system is  always  being in  a pure state.
  Sometimes this system is
 called Tavis-Cummings model \cite{tavis} or Dicke model \cite{dick} or two-atom JCM.
For this system  we make a comparative study to the atomic
 inversions, the entropy squeezing, the atomic variances and the von Neumann entropy.
 We treat two cases,
namely, symmetric (two identical atoms) and asymmetric (two
non-identical atoms) cases. The  investigation will be restricted
 to the single-atom case, where  there are difficulties
to deal with the entropy squeezing of the compound case, as we
argue in section 2. We should stress that the behavior of the
individual atoms in the TJCM is generally different from that in
the JCM, where the distribution of  energy among the parties in
the bipartite is completely different from that in the tripartite.
  Such type of investigation is
motivated by the importance of the TJCM  in the literatures, e.g.
\cite{ros1,{igo},{tess},{ros5},{nf},{nff}}. The atomic inversions
of the TJCM  have taken much interest \cite{ros1,{nf},{nff}} since
they  exhibit different shapes of the revival-collapse phenomenon
(RCP). The importance of the TJCM has  increased as a result of
the progress in the quantum information \cite{infor,{wott}}. In
this respect, the entanglement for the TJCM with the initial
coherent state \cite{tess}, the binomial state \cite{ros5} and the
superposition displaced Fock state \cite{nff} has been
investigated. Moreover, various schemes have been proposed for the
TJCM, e.g., \cite{janos}.
 This information indicates that the investigation presented in
this paper is important in its own right. Additionally, we
obtained many of interesting results.
 For instance, as is well known for the JCM that  the
entropy squeezing always gives better information on the atomic
system than the atomic variances \cite{fang,{obo}}. In  this paper
we show that this is not always correct, where in some cases they
can give identical behaviors. Also when the number of the qubits
 in the quantum system is increased the amounts of the nonclassical
effects occurred in the entropy squeezing decrease. More
precisely, we show that  the amounts of the nonclassical effects
in the entropy squeezing associated with the JCM are greater than
those with the  TJCM.
 Most importantly, we show that  the entropy
squeezing can generally give information on the corresponding von
Neumann entropy.   Also the amounts of the nonclassical effects
obtained form the asymmetric case are greater than those obtained
from the symmetric one. These results are  valid for both $l=1$
and $l=2$. Finally, throughout the paper the phrase "nonclassical
effects" means that the entropy squeezing and/or the atomic
variance include negative values.

The paper is prepared in the following order. In section 2 we give
the basic equations and relations for the system under
consideration. In section 3 we investigate the atomic
 inversions, the entropy squeezing, the atomic variances and the von Neumann entropy.
 The main conclusions are summarized in section 4.

\section{Basic equations and relations}
In this section we give a definition for the atomic squeezing in
terms of  the entropic information and the atomic variances. We
develop the Hamiltonian and the wavefunction for a two two-level
atoms multi-photon  JCM. Also  we deduce the expressions of the
entropy squeezing and the von Neumann entropy.

 As is well known that for
the $j$th atom the Pauli spin operators $\hat{\sigma}^{(j)}_x,
\hat{\sigma}^{(j)}_y$ and $\hat{\sigma}^{(j)}_z$ determine the
real, the imaginary parts of the complex dipole moment and the
energy of  the atom, respectively. This set of operators satisfy
the following commutation rule:
\begin{equation}
\left[\hat{\sigma}^{(j)}_x, \hat{\sigma}^{(j)}_y\right]=
2i\hat{\sigma}^{(j)}_z.
 \label{en1}
\end{equation}
The  Heisenberg uncertainty relation (HUR) associated with
(\ref{en1}) is
\begin{equation}
\langle\left(\triangle\hat{\sigma}^{(j)}_x\right)^2\rangle
\langle\left(\triangle\hat{\sigma}^{(j)}_y\right)^2\rangle\geq
|\langle\hat{\sigma}^{(j)}_z\rangle|^2.
 \label{en1n}
\end{equation}
From (\ref{en1n}) the atomic system has reduced fluctuations (,
i.e. squeezing) in the $\hat{\sigma}^{(j)}_x$ or in the
$\hat{\sigma}^{(j)}_y$ if
\begin{equation}
F_k^{(j)}=\langle\left(\triangle\hat{\sigma}^{(j)}_k\right)^2\rangle-
|\langle\hat{\sigma}^{(j)}_z\rangle|<0,\qquad k=x, y. \label{en2}
\end{equation}
The inequality (\ref{en1n})  is a state dependent and it is
 trivially satisfied for any  atomic state having
 $\langle\hat{\sigma}^{(j)}_z\rangle=0$ \cite{fang}.
In this case (\ref{en2}) fails  to provide any useful information
on the atomic system. This difficulty has been overcome  using the
entropic uncertainty relation (EUR) \cite{hirs,{maa}}. An optimal
EUR for a set of $N+1$ complementary observable with different
eigenvectors in an even $N$-dimensional Hilbert space can be
evaluated through the inequality \cite{sanc}:

\begin{equation}
\sum\limits_{k=1}^{N+1}H(\hat{\sigma}^{(j)}_k)\geq \frac{N}{2}{\rm
ln}(\frac{N}{2}) + (\frac{N}{2}+1){\rm ln}(\frac{N}{2}+1),
\label{en3}
\end{equation}
where $H(\hat{\sigma}^{(j)}_k)$ is the information entropy
associated with the variable $\hat{\sigma}^{(j)}_k$. The quantity
$H(\hat{\sigma}^{(j)}_k)$ can be described as follows:
 For an arbitrary atomic  system
described by the density matrix $\hat{\rho}$, the probability
distribution of $N$ possible outcome of measurements of the
operator $\hat{\sigma}^{(j)}_k$ is
\begin{equation}
P_{j'}(\hat{\sigma}^{(j)}_k)=\langle
\psi_{kj'}|\hat{\rho}|\psi_{kj'}\rangle,\qquad j'=1,2,...,N,
 \label{en4}
\end{equation}
where $|\psi_{kj'}\rangle$ are the eigenstates of
$\hat{\sigma}^{(j)}_k$. In this case the associated information
entropy is:
\begin{equation}
H(\hat{\sigma}^{(j)}_k)=-\sum\limits_{j'=1}^{N}P_{j'}(\hat{\sigma}^{(j)}_k)
{\rm ln}P_{j'}(\hat{\sigma}^{(j)}_k).
 \label{en6}
\end{equation}
For the single-atom JCM we have $N=2$ and then $0\leq
H(\hat{\sigma}^{(j)}_k)\leq {\rm ln}2$, where $k=x,y,z$. For this
case the inequality (\ref{en3}) takes the form:
\begin{equation}
H(\hat{\sigma}^{(j)}_x) +H(\hat{\sigma}^{(j)}_y)\geq {\rm ln}4-
H(\hat{\sigma}^{(j)}_z).
 \label{en7}
\end{equation}
From  (\ref{en7}) the components $\hat{\sigma}^{(j)}_k(k\equiv
x,y)$ are said to be squeezed with respect to  the information
entropy if one  or both of them satisfy the condition \cite{fang}:

\begin{equation}
E_k^{(j)}=\delta H(\hat{\sigma}^{(j)}_k)-\frac{2}{\sqrt{ \delta
H(\hat{\sigma}^{(j)}_z})}<0,\qquad k=x,y,
 \label{en8}
\end{equation}
where $\delta H(\hat{\sigma}^{(j)}_k)=\exp[
H(\hat{\sigma}^{(j)}_k)]$. It is worth mentioning that  $\delta
H(\hat{\sigma}^{(j)}_k)=1 \quad (\delta
H(\hat{\sigma}^{(j)}_k)=2)$
 corresponds to the atom being in a pure (mixed) state.
As we mentioned in the Introduction that throughout the paper the
phrase "nonclassical effects" means that $E_{k}^{(j)}<0$ or
$F_{k}^{(j)}<0$.
 Furthermore, the
  optimal nonclassical entropy squeezing
$E_{k}^{(j)}\simeq -0.4140$  is associated with the eigenstates

\begin{equation}\label{eigen}
|\phi^{(x)}_\pm\rangle=\frac{1}{\sqrt{2}}(|+\rangle\pm|-\rangle),
\quad |\phi^{(y)}_\pm\rangle=\frac{1}{\sqrt{2}}(|+\rangle \pm
i|-\rangle),
\end{equation}
where $|+\rangle$ and $|-\rangle$ denote the excited and the
ground atomic states; and the superscripts $(x)$ and $(y)$ mean
that the nonclassical effects only  occur in $E_x(.)$
 and $E_y(.)$, respectively. Throughout the paper we
only study the entropy squeezing for the single-atom case
 since the EUR (\ref{en3}) is not relevant for  $N
(\geq 2)$ atomic system, where, e.g., for the TJCM  there are
degenerate eigenvalues for the compound spin operators
$\hat{\sigma}^{(1)}_x+\hat{\sigma}^{(2)}_x$ and
$\hat{\sigma}^{(1)}_y+\hat{\sigma}^{(2)}_y$ \cite{fang}.

The Hamiltonian describing the two two-level atoms  interacting
with the single-mode electromagnetic field through multi-photon
transition, namely, two atoms Jaynes-Cummings model (TJCM), in the
rotating wave approximation takes the form, e.g.,
\cite{ros1,{igo},{tess},{ros5},{nf},{nff}}:

\begin{eqnarray}
\begin{array}{lr}
\frac{\hat{H}}{\hbar}=\hat{H}_0+\hat{H}_I,\\
\\
\hat{H}_0= \omega\hat{a}^{\dagger}\hat{a}+\frac{1}{2}
\omega_a(\hat{\sigma}_z^{(1)}+\hat{\sigma}_z^{(2)}),\quad
\hat{H}_I=\sum\limits_{j=1}^2 \lambda_j
(\hat{a}^l\hat{\sigma}_+^{(j)} + \hat{a}^{\dagger l
}\hat{\sigma}_-^{(j)}),
 \label{6}
 \end{array}
\end{eqnarray}
where $\hat{H}_0$ and $\hat{H}_I$ are the free and the interaction
parts of the Hamiltonian, $\hat{\sigma}_\pm^{(j)}$ and
$\hat{\sigma}_z^{(j)}$ are the Pauli spin operators of the $j$th
atom; $\hat{a}\quad (\hat{a}^{\dagger})$ is the annihilation
(creation) operator denoting  the cavity mode, $\omega$ and
$\omega_a$ are the frequencies of the cavity mode and the atomic
systems (we  assume that the two atoms have the same frequency),
$\lambda_j$ is the atom-field coupling constant of the $j$th atom
and $l$ is the transition parameter. From the Hamiltonian
(\ref{6}) it is evident that the atoms do not interact directly,
but only through the common radiation field.  Throughout the
investigation we mainly deal with the ratio
$g=\lambda_2/\lambda_1$, when $g=1 \quad (g\neq 1)$ it is called
symmetric (asymmetric) case. Additionally, we assume that
$\omega_a=2l\omega$ (, i.e. the exact resonance case) and the two
atoms and the field are initially prepared in the excited atomic
states  and coherent state $|\alpha\rangle$, respectively. Under
these initial conditions,  the dynamical state of the system can
be easily evaluated as \cite{nf}:
\begin{eqnarray}
\begin{array}{lr}
|\Psi(T)\rangle=
\sum\limits_{n=0}^{\infty}C_n\left[X_1(T,n)|+,+,n\rangle
+iX_2(T,n)|+,-,n+l\rangle\right.
\\
\\
\left. +iX_3(T,n)|-,+,n+l\rangle +X_4(T,n)|-,-,n+2l\rangle
\right], \label{10}
\end{array}
\end{eqnarray}
where  $T=\lambda_1 t$ is the scaled time and
$C_n=\frac{\alpha^n}{\sqrt{n!}}\exp(-\frac{\alpha^2}{2})$ with the
real amplitude $\alpha$. The explicit general forms for the real
dynamical coefficients $X_j(T,n)$ are given, e.g., in \cite{nf}.
Nevertheless,  for reasons will be made clear shortly we give the
forms of these coefficients for the case $(l,g)=(1,1)$ \cite{nff}
as:
\begin{eqnarray}
\begin{array}{lr}
X_1(T,n)= \frac{1}{2n+3}[(n+1)\cos(T\theta_n)+(n+2)],\\
\\
X_2(T,n)=X_3(T,n)=- \frac{\sqrt{n+1}}{\theta_n}\sin(T\theta_n),\\
\\
X_4(T,n)=\frac{\sqrt{(n+1)(n+2)}}{2n+3}[\cos(T\theta_n)-1], \quad
\theta_n=\sqrt{4n+6}. \label{coef}
\end{array}
\end{eqnarray}

Now we are in a position to calculate the single-atom entropy
squeezing, the atomic variances and the von Neumann entropy for
the first atom. In doing so we assume that $A_1, A_2$ and $f$
denote the first atom, the second atom and the radiation field,
respectively. The density matrix of the whole system is
$\hat{\rho}_{A_1A_2f}(T)=|\Psi(T)\rangle \langle \Psi(T)|$, where
$|\Psi(T)\rangle$ is given by (\ref{10}). As
 we treat the evolution of the  single-atom case we have to
 trace out the remaining part of the density matrix.
 For instance, the density matrix of
the first atom can be obtained as:

\begin{eqnarray}
\begin{array}{lr}
\hat{\rho}_{A_1}(T)={\rm Tr}_{A_2f}\hat{\rho}_{A_1A_2f}(T),
\\
\\
=\sum\limits_{n=0}^{\infty}\left[Q_1(n,n)|+\rangle\langle +|+
Q_2(n,n)|-\rangle\langle -|+ Q_3(n,n+l)|+\rangle\langle -|+
Q^{*}_3(n,n+l)|-\rangle\langle +|\right],
 \label{I3}
 \end{array}
\end{eqnarray}
where
\begin{eqnarray}
\begin{array}{lr}
Q_1(n,n)=C_{n,n}[X^2_1(T,n)+X^2_2(T,n)],\\
\\
Q_2(n,n)=C_{n,n}[X^2_3(T,n)+X^2_4(T,n)],\\
\\
Q_3(n,n+l)=iC_{n+l,n}[X_2(T,n+l)X_4(T,n)-X_3(T,n)X_1(T,n+l)]
 \label{I3I}
\end{array}
\end{eqnarray}
and $C_{n,m}=C_{n}C_{m}$. By means of (\ref{I3}) one can deduce
the followings:
\begin{eqnarray}
\begin{array}{lr}
\langle \sigma^{(1)}_z(T)\rangle= \sum\limits_{n=0}^{\infty}[
Q_1(n,n)-Q_2(n,n)],\\
\\
\langle \sigma^{(1)}_x(T)\rangle=2{\rm Re}
\sum\limits_{n=0}^{\infty}Q_3(n,n+l)=0,\\
\\
\langle \sigma^{(1)}_y(T)\rangle=2{\rm Im}
\sum\limits_{n=0}^{\infty}Q_3(n,n+l),
 \label{I3iI}
\end{array}
\end{eqnarray}
where ${\rm Re}$ and ${\rm Im}$ stand for the real and imaginary
parts of the complex quantity. From (\ref{en4}), (\ref{en6}) and
(\ref{I3}) one can obtain the information entropies of the atomic
operators $\sigma^{(1)}_x, \sigma^{(1)}_y, \sigma^{(1)}_z$ as

\begin{eqnarray}
\begin{array}{lr}
H(\hat{\sigma}^{(1)}_x)=-\frac{1}{2}[1+\langle
\sigma^{(1)}_x(T)\rangle]{\rm ln} [\frac{1}{2}+\frac{1}{2}\langle
\sigma^{(1)}_x(T)\rangle] -\frac{1}{2}[1-\langle
\sigma^{(1)}_x(T)\rangle]{\rm ln} [\frac{1}{2}-\frac{1}{2}\langle
\sigma^{(1)}_x(T)\rangle]={\rm ln}2,\\
\\
H(\hat{\sigma}^{(1)}_y)=-\frac{1}{2}[1+\langle
\sigma^{(1)}_y(T)\rangle]{\rm ln} [\frac{1}{2}+\frac{1}{2}\langle
\sigma^{(1)}_y(T)\rangle] -\frac{1}{2}[1-\langle
\sigma^{(1)}_y(T)\rangle]{\rm ln} [\frac{1}{2}-\frac{1}{2}\langle
\sigma^{(1)}_y(T)\rangle],\\
\\
H(\hat{\sigma}^{(1)}_z)=- \frac{1}{2}[1+\langle
\sigma^{(1)}_z(T)\rangle]{\rm ln} [\frac{1}{2}+\frac{1}{2}\langle
\sigma^{(1)}_z(T)\rangle] - \frac{1}{2}[1-\langle
\sigma^{(1)}_z(T)\rangle]{\rm ln} [\frac{1}{2}-\frac{1}{2}\langle
\sigma^{(1)}_z(T)\rangle].
 \label{IaI}
\end{array}
\end{eqnarray}
We close this section by writing the expression  of the
single-atom von Neumann entropy as:
\begin{equation}\label{von1}
\gamma(T)=-{\rm Tr}[\hat{\rho}_{A_1}(T){\rm
ln}\hat{\rho}_{A_1}(T)]
    = -\mu_{-}(T){\rm
ln}\mu_{-}(T)-\mu_{+}(T){\rm ln}\mu_{+}(T),
\end{equation}
where  $\mu_{\pm}$ are the  eigenvalues of the density matrix
$\hat{\rho}_{A_1}(T)$, which can be easily evaluated from
(\ref{I3}) as:
\begin{equation}\label{von2}
\mu_{\pm}=\frac{1}{2}\sum\limits_{n=0}^{\infty}[Q_1(n,n)+Q_2(n,n)]\pm
\frac{1}{2}\sqrt{\{\sum\limits_{n=0}^{\infty}[Q_1(n,n)-Q_2(n,n)]
\}^2+4|\sum\limits_{n=0}^{\infty}Q_3(n,n+l)|^2}.
\end{equation}
For the future purpose, by means of  (\ref{I3iI}) the eigenvalues
(\ref{von2}) can be re-expressed as:
\begin{equation}\label{von3}
 \mu_{\pm}=\frac{1}{2}\pm \frac{1}{2}\sqrt{\langle
\sigma^{(1)}_x(T)\rangle^2+\langle
\sigma^{(1)}_y(T)\rangle^2+\langle \sigma^{(1)}_z(T)\rangle^2}.
\end{equation}

 It is worth mentioning that the relations
related to the second atom can be obtained from those of the first
one by using the interchange $X_2(T,n)\leftrightarrow X_3(T,n)$.
In the following section we use the relations obtained above to
investigate the single-atom entropy squeezing, the atomic
variances, the von Neumann entropy and the atomic inversions for
the system under consideration.
\begin{figure}
  \vspace{0cm}
    \includegraphics[width=0.86\linewidth]{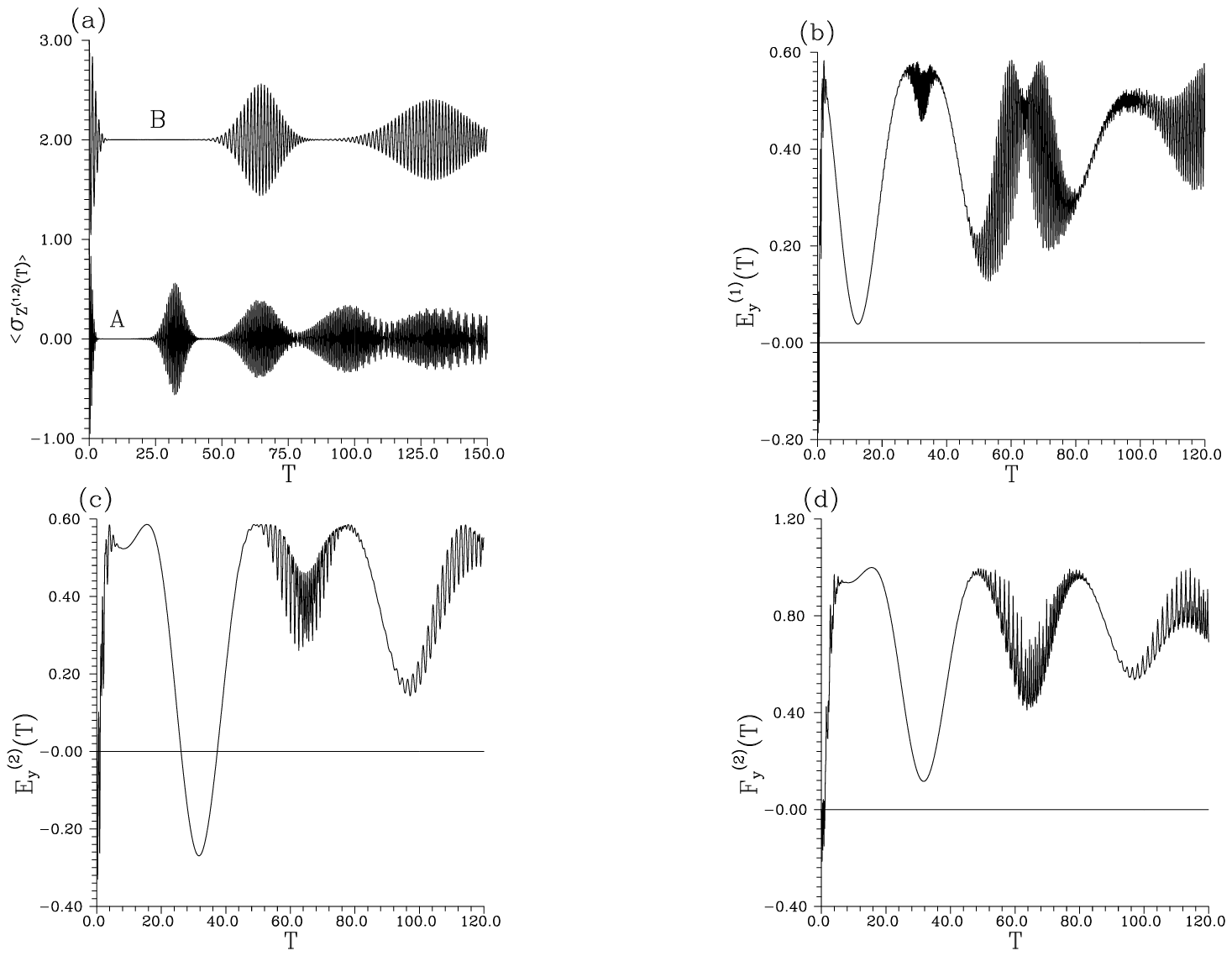}
\caption{ The evolution of the atomic inversions, the entropy
squeezing and the atomic variance as indicated
 for $(\alpha,g,l)=(5,0.5,1)$. In (a) the curves A and B are given
 for
  $\langle \hat{\sigma}^{(1)}_z(T)\rangle$ and
$\langle \hat{\sigma}^{(2)}_z(T)\rangle+2$, respectively. The
straight lines in  (b)--(d) are given to show the nonclassical
effects bounds.}
\end{figure}

\section{ Discussion of the results}
In this section we investigate  the single-atom atomic inversions,
the entropy squeezing, the
 atomic variances and the von Neumann entropy for the system under consideration.
From (\ref{en8}), (\ref{I3iI}) and (\ref{IaI}) one can prove for
the $x$-component entropy squeezing  that
\begin{equation}
E_x^{(j)}(T)=2\left[1-\frac{1}{\sqrt{ \delta
H(\hat{\sigma}^{(j)}_z})}\right]\geq 0,
 \label{enn8}
\end{equation}
where $\frac{1}{\sqrt{ \delta H(\hat{\sigma}^{(j)}_z})}\leq 1$.
This means that $E_x^{(j)}$ cannot exhibit nonclassical effects.
  Similar arguments show that
$F_x^{(j)}\geq 0$. Thus throughout the discussion we focus the
attention on the $y$-component of both the entropy squeezing and
 the atomic variances. Furthermore, for a weak initial field
intensity, i.e. $\alpha\leq 1$, we have found that $E_y^{(j)}$ and
$F_y^{(j)}$ cannot exhibit nonclassical effects. This is because
 the coherent state $|\alpha\rangle$ tends to the vacuum
state $|0\rangle$ and/or the Fock state, which manifest themselves
as a periodic behavior in these quantities.  Moreover, we have
noted that the entropy squeezing cannot exhibit nonclassical
effects for $l>2$.  When the initial field intensity is strong the
atomic inversion of the  system exhibits the RCP, which is
connected with the occurrence of the nonclassical effects in the
system. For instance,  the JCM generates the Schr\"{o}dinger-cat
states  at one-half of the revival time \cite{fais}, however, the
symmetric (asymmetric) TJCM generates asymmetric (symmetric) cat
states at the quarter of the revival time \cite{nff}.
Additionally, for the JCM the entropy squeezing exhibits
nonclassical effects only in the course of the collapse regions of
the corresponding atomic inversion, however, the atomic variance
fails to give any useful information \cite{fang}. As a result of
these facts we'll study the evolution of the atomic inversions for
the TJCM, too. As we mentioned in the Introduction that the
behavior of the individual atoms in the TJCM is generally
different from that in the JCM, where the latter (the former)
includes one (two) interaction mechanism(s).

Now we start the investigation with  the asymmetric case for the
single-photon transition mechanism. For this case we plot the
atomic inversions, the entropy squeezing and the atomic variance
in Figs. 1 as indicated for the given values of the interaction
parameters. In these figures we take $\lambda_1=2\lambda_2$, i.e.
the interaction between the field and the first atom is two times
stronger than that with the second atom. This is manifested in the
evolution of the atomic inversions, where the collapse regions  in
 $\langle \hat{\sigma}^{(2)}_z(T)\rangle$ are two times greater
than those in  $\langle \hat{\sigma}^{(1)}_z(T)\rangle$ (see Fig.
1(a)). This means that the rate of energy interchange  between the
radiation field and the first atom is two times faster than that
with the second atom. The behaviors  of the atomic inversions are
roughly connected with
 the evolution of the entropy squeezing and the atomic
variance (see Fig. 1(b)--(d)). For instance, $E_{y}^{(1)}(T)$ and
$E_{y}^{(2)}(T)$ exhibit nonclassical effects immediately after
switching on the interaction. At this stage the atomic inversions
provide their zero  revival patterns. As the interaction proceeds
 the nonclassicality completely disappears  in
$E_{y}^{(1)}(T)$, however,  $E_{y}^{(2)}(T)$ provides its maximum
value approximately at one-half of the revival time in the
$\langle \hat{\sigma}^{(2)}_z(T)\rangle$. The comparison between
Fig. 1(b) and Fig. 1(c) shows that the amounts of the nonclassical
effects exhibited in  $E_{y}^{(2)}(T)$ are much greater than those
in  $E_{y}^{(1)}(T)$. These amounts  can be increased by
increasing the value of  $\alpha$.  We have  checked this fact.
 Additionally, the behaviors of
$E_{y}^{(1)}(T)$ and $E_{y}^{(2)}(T)$
  in the Fig. 1(b) and (c) can be reversed if one
takes $\lambda_2=2\lambda_1$ and $T=t\lambda_2$. On the other
hand, we have noted that  $F_y^{(1)}(T)$ and $F_y^{(2)}(T)$
provide similar behaviors. This indicates that the entropy
squeezing is more sensitive to the interaction mechanisms in the
compound system
 than the atomic variance. We have plotted only
$F_y^{(2)}(T)$ in Fig. 1(d). From this figure we can see that
$F_y^{(2)}(T)$ exhibits nonclassical effects only after switching
on the interaction. Comparison between Fig. 1(c) and Fig. 1(d)
confirms that the entropy squeezing  gives better information on
the atomic system than the atomic variance \cite{fang}. We'll show
below that this statement is not always correct.
\begin{figure}
  \vspace{0cm}
    \includegraphics[width=0.86\linewidth]{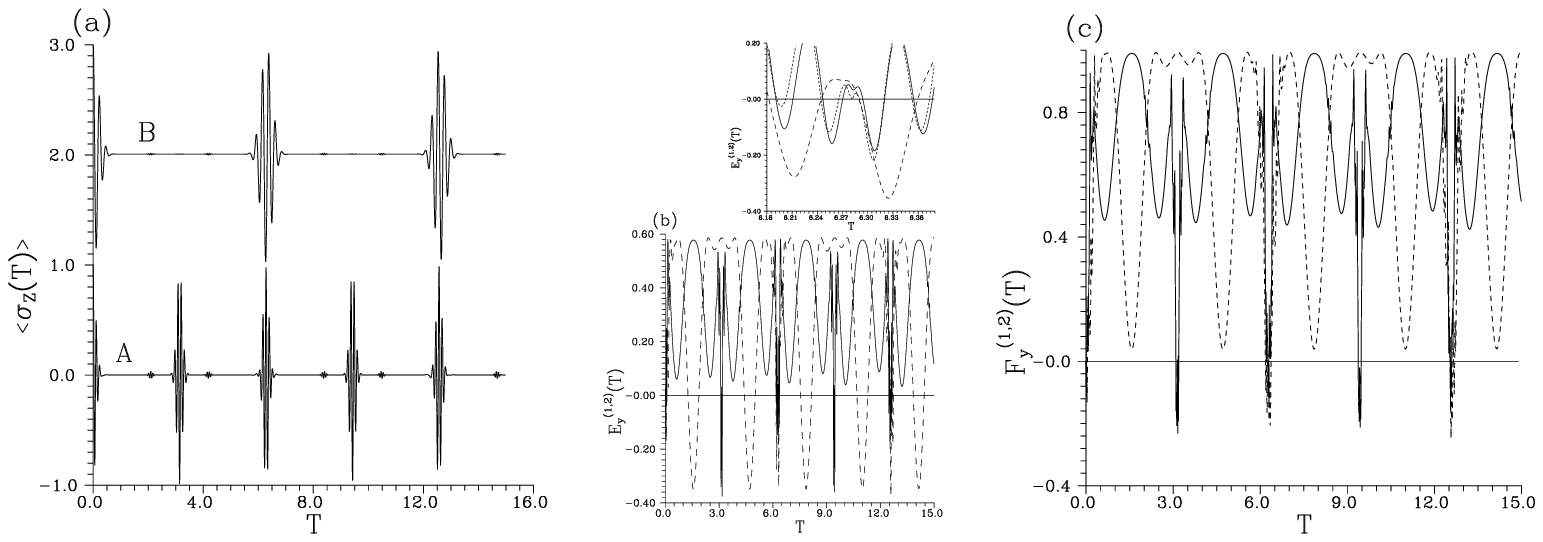}
\caption{ The evolution of the atomic inversions, the entropy
squeezing and the atomic variance as indicated in the
 figures for $(\alpha,g,l)=(5,0.5,2)$.
 In (a) the curves A and B are given for
  $\langle \hat{\sigma}^{(1)}_z(T)\rangle$ and
$\langle \hat{\sigma}^{(2)}_z(T)\rangle+2$, respectively. In (b)
and (c) the solid and dashed curves are given for the first and
second atom, respectively. The inset in (b) is given to show the
behavior of the entropy squeezing through a very short period
around $T=2\pi$. The short-dashed curve in this inset is given for
$E_y$ of the case $g=1$. The straight lines in (b)--(c) are given
to show the nonclassical bounds. }
\end{figure}

Now we draw the attention to the two-photon transition case, i.e.
$l=2$, which is plotted in  Figs. 2 for the given values of the
interaction parameters. From Fig. 2(a)
$\langle\hat{\sigma}^{(1)}_z(T)\rangle$ and $\langle
\hat{\sigma}^{(2)}_z(T)\rangle$  exhibit periodic, systematic, and
compact RCP with periods $\pi$ and $2\pi$,
 respectively. This behavior  is connecting with  the two-photon nature
 of the system.
Also one can observe the occurrence of the subsidiary revivals in
the evolution of $\langle\hat{\sigma}^{(1)}_z(T)\rangle$, i.e.
each revival pattern is followed by a subsidiary one (see the
curve A in Fig. 2(a)).
 It is worth reminding  that the subsidiary revivals
have been observed also for the two-mode single-photon JCM
\cite{card}. On the other hand, from the solid curve in the Fig.
2(b) one can observe that  $E_y^{(1)}(T)$  periodically (with
period $\pi$) exhibits  nonclassical effects  in the course of the
revival patterns in $\langle\hat{\sigma}^{(1)}_z(T)\rangle$.
Moreover, the amounts of the nonclassical effects occurred in
$E_y^{(1)}(T)$ around $T=s\pi$ are more pronounced than those
around $T=s'\pi$, where $s$ and $s'$ are odd and even integers,
respectively. Also we have found $E_y^{(1)}(T=\widetilde{s}\pi)
\simeq 0$, where $\widetilde{s}$ is integer. For instance, this is
obvious from the inset in Fig. 2(b) for a short period around
$T=2\pi$. Next, the evolution of $E_y^{(2)}(T)$ exhibits
periodical compound behavior with period $2\pi$ (see the dashed
curve in Fig. 2(b)). More illustratively, $E_y^{(2)}(T)$ exhibits
long-lived nonclassical effect at $T\simeq \pi/2, 3\pi/2$ and
instantaneous nonclassical effect around $T=2\pi$. The comparison
between the curve B in the Fig. 2(a) and the dashed curve in the
Fig. 2(b) shows that $E_y^{(2)}(T)$ provides nonclassical effects
in the course of both of the collapse regions and the revival
patterns of $\langle \hat{\sigma}^{(2)}_z(T)\rangle$.
 From the dashed-curve in the inset in the
 Fig. 2(b), one can realize that $E_y^{(2)}(T=2\widetilde{s}\pi)\simeq 0$.
 We'll return to this point shortly.
  From the above investigation it is obvious for the case $l=2$ that
  there is no clear relationship between the occurrence of the
 RCP in the atomic inversions and  the
nonclassical effects in the entropy squeezing. Now we draw the
attention to $F_y^{(1,2)}(T)$. Generally,  the
 behaviors of  $F_y^{(1,2)}(T)$ are similar to those
  of  $E_y^{(1,2)}(T)$ (compare Fig. 2(b) and Fig. 2(c)).
Nevertheless, $F_y^{(1,2)}(T)$ provide only instantaneous
nonclassical effects with amounts  smaller than those exhibited in
 $E_y^{(1,2)}(T)$.

\begin{figure}
  \vspace{0cm}
    \includegraphics[width=0.86\linewidth]{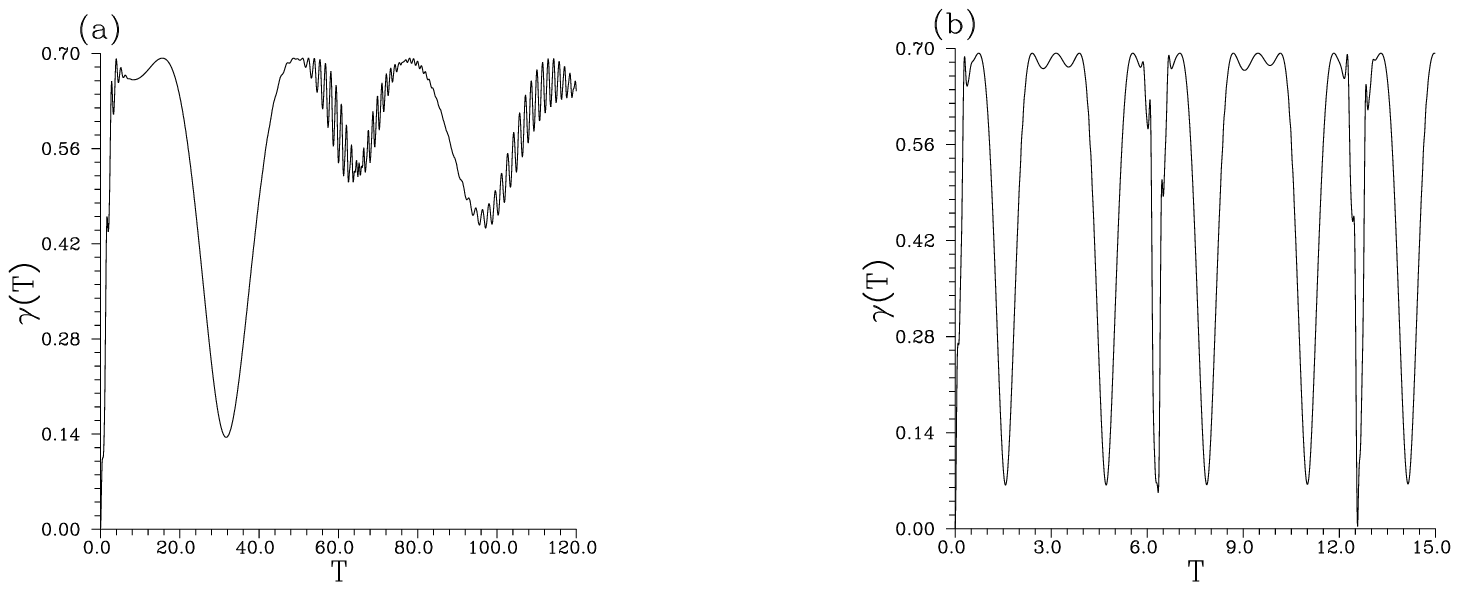}
 \caption{
 The evolution of the von Neumann entropy for the second
 atom for $(\alpha,g)=(5,0.5)$ when $l=1$ (a) and $l=2$
(b). }
\end{figure}
We conclude this part by showing an important fact: the entropy
squeezing can give information on the corresponding von Neumann
entropy. To clarify  this point we plot--as an example--$\gamma
(T)$ for the second atom in the Fig. 3(a) and (b) when $l=1$ and
$2$, respectively. The comparison between these figures and the
corresponding $E_y^{(2)}$  in Figs. 1 and 2 is instructive. Of
course these quantities include different scales, e.g. $0\leq
\gamma(T)\leq {\rm ln}2$ and $-0.4\preceq E_y^{(j)}(T)\preceq 0.6$
(see Figs. 1-2). The limitations of the $E_y^{(j)}(T)$ have been
numerically obtained. From Figs. 1--3 one can realize when
$(E_y^{(2)},\gamma)\simeq (-0.4,0)$ or $ (0.6,0.639)$ the
bipartite (, i.e. $A_2,fA_1$) is disentangled or maximally
entangled. In this respect, the entropy squeezing can play two
roles, one for squeezing and the other for entanglement. For the
latter, it can be interpreted as follows: Apart from $T>0$ for
$E_j<0 \quad (>0)$ the bipartite will be close to the disentangled
(entangled) form till $E_j=-0.41\quad (0.6)$ the bipartite will be
in a complete disentangled (maximally entangled) states.
Additionally, the bipartite is disentangled when $E_j(T)$ has an
inclined point at $E_j(T)=0$. The inclined point means that
$E_j(T)$ changes its behavior   around this point, i.e. the
function changes its behavior  from an increasing function to a
decreasing one or vice versa around $E_j(T)=0$. This is quite
obvious from the inset in Fig. 2(b). Also from this inset one can
realize that the tripartite is
 periodically (with period $2\pi$) in an approximate disentangled form, e.g.
$\hat{\rho}(T=2\widetilde{s}\pi)
\simeq\hat{\rho}_{A_1}\bigotimes\hat{\rho}_{A_2}\bigotimes\hat{\rho}_{f}$,
however, it is difficult to find the asymptotic  forms for these
 density matrices. Nevertheless,
for the symmetric case these forms have been already derived in
\cite{nff} and confirmed by the short-dashed curve in the inset in
Fig. 2(b), which involves an explicit inclined point at $T=2\pi$.
Now, the reason why  the entropy squeezing $E_y^{(j)}(T)$ and the
von Neumman entropy have similar behaviors is that both of them
are functions in the $\langle\sigma_z^{(2)}(T)\rangle$ and
$\langle\sigma_y^{(2)}(T)\rangle$ (c.f. (\ref{en8}),
(\ref{IaI})--(\ref{von3})). Also from these equations, for
$(\langle\sigma_z^{(2)}(T)\rangle,\langle\sigma_y^{(2)}(T)\rangle
=(0,\pm 1)$ and $(\pm 1,0)$ one can prove
$(E_y^{(2)},\gamma)\simeq (-0.4,0)$ and $(0,0)$, respectively.
 These results are
in a good agreement with the above discussion. We have to stress
that this relationship between the von Neumann and the entropy
squeezing is universal (for the JCM and TJCM) provided that the
atoms are initially prepared in the excited or ground states. We
have checked this fact. The final remark: the inverse situation is
not correct. More illustratively, the von Neumann entropy cannot
be used to give information on the nonclassicality in the entropy
squeezing. For instance,  when $\gamma(T)\simeq 0$ this may be
corresponding  to $E_j=-0.41$ or $E_j=0$, as we have shown above.
\begin{figure}
  \vspace{0cm}
    \includegraphics[width=0.86\linewidth]{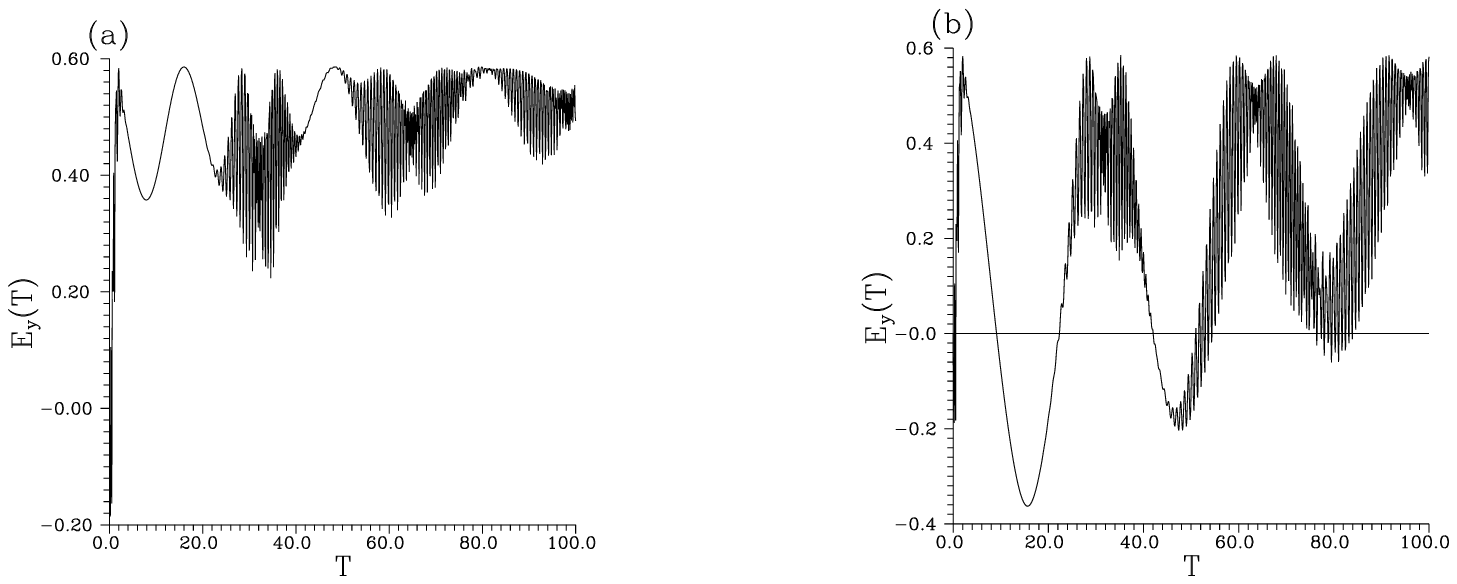}
\caption{
 The evolution of the $E_y(T)$ for the TJCM (with $g=1$) (a)
 and the JCM (b) when $(\alpha,l)=(5,1)$.
 The straight line in  (b) is  given to show the nonclassical effects bound. }
\end{figure}

 We close this section by discussing the symmetric case, in particular, $(g,l)=(1,1)$.
 In this case the energy
interchanged between the field and the atoms is equally
distributed between the two atoms. Therefore, the evolution of the
different quantities related to the individual  atoms are
identical. For this case  the atomic inversion  has an identical
behavior with that of the JCM (see the curve A Fig. 1(a)). In
spite of this fact the evolution of  $E_y(T)$ of the JCM and of
the TJCM are completely different (see Figs. 4). From Fig. 4(a)
one can observe that  $E_y(T)$ exhibits nonclassical effects only
after switching on the interaction, however, this is not the case
for the JCM and/or Fig. 4(b). Moreover, the comparison between
these figures  shows that the amounts of the nonclassical effects
associated with  $E_y(T)$ of the JCM are much greater than those
exhibited for the TJCM. This conclusion is also valid for the
asymmetric case (compare Fig. 1(c) to Fig. 4(b)).
 This means when the number of the qubits
 in the quantum system is increased, e.g. Ising model \cite{isi}, the
amounts of the nonclassical effects in $E_k(T)$ of the individual
qubits  decrease. We can analytically explain
 this  fact for the TJCM as follows:   $E_y$
depends  on $\langle\sigma_z(T)\rangle$ and
$\langle\sigma_y(T)\rangle$, however, for the JCM and the
symmetric TJCM the quantity $\langle\sigma_z(T)\rangle$ has a
quite similar behavior for both of them, as we mentioned above.
Thus $\langle\sigma_y(T)\rangle$ plays the essential role in the
evolution of the entropy squeezing.
 When $\alpha>>1$ the probability distribution $P(n)=C_n^2$ is Piossonian
and has the main contribution around $n\simeq \alpha^2$. In this
case  the harmonic approximation  can be applied to (\ref{coef})
(, i.e. $\epsilon/n\rightarrow 1$, where $\epsilon$ is an
arbitrary c-number) and hence the quantity
$\langle\sigma_y(T)\rangle$ in (\ref{I3iI}) can be simplified as:
\begin{equation}\label{sgy1}
\langle\sigma_y(T)\rangle\simeq\sum\limits_{n=0}^{\infty}C_{n,n+1}
\{\frac{1}{2}\sin[T (\theta_n-\theta_{n+1})]+
\sin[T(\frac{\theta_n+\theta_{n+1}}{2})]
\cos[T(\frac{\theta_n-\theta_{n+1}}{2})]\}.
\end{equation}
Nevertheless, for the JCM we have
\begin{equation}\label{sgy2}
\langle\sigma_y(T)\rangle=2\sum\limits_{n=0}^{\infty}C_{n,n+1}
\cos(T\sqrt{n+2})\sin(T\sqrt{n+1}).
\end{equation}
The comparison between the two above expressions leads to that
generally the amplitudes of the $\langle\sigma_y(T)\rangle$ of the
TJCM is one-half of that of the JCM.  We have numerically checked
this fact. This completes the explanation. Similar conclusions
have been also realized for the case $l=2$. Finally, we have found
for the symmetric case  that $F_y(T)$ and $E_y(T)$ provide similar
behaviors. This indicates that the entropy squeezing does not
always give better information than the atomic variances.

\section{Conclusion }
In this paper we have investigated a system of two two-level atoms
interacting with a single-mode quantized electromagnetic field in
a lossless resonant cavity via $l$-photon-transition mechanism.
The  atoms and the field are initially prepared in the excited
atomic states and the coherent state, respectively. We have found
 the partial density matrix for an individual atom
      by tracing the state over the variables of the other atom and the field.
Two cases have been treated, namely, the symmetric and the
asymmetric  cases. The investigations have been focused on  the
entropy squeezing, the atomic variances, the von Neumann entropy
and the atomic inversions for the pseudo-spin components of the
individual atom. We have shown that the values of the transition
parameter $l$, the ratio $g$ and the intensity $\alpha$ are
important for the occurrence of the nonclassical effects in the
entropy squeezing. We have numerically shown that the system
cannot provide nonclassical effects in $E^{(j)}_k$ for $l>2$. The
amounts of the nonclassical effects in $E^{(j)}_k$ obtained from
the TJCM are smaller than those obtained from the JCM. In other
words, when the number of the parties in the quantum system is
increased the amounts of the nonclassical effects involved in the
entropy squeezing decrease. The nonclassical effects in
$E^{(j)}_k$ obtained from the asymmetric case are greater than
those obtained from the symmetric one.
 Also we have shown that
 the entropy squeezing does not always give better information than the atomic
variances. Most importantly, we have shown that  the entropy
squeezing can generally give information on the corresponding von
Neumann entropy.    These results are verified  for both of the
cases $l=1$ and $l=2$.
  Additionally, there is no clear relationship between the
occurrence of the RCP in the atomic inversions and the
nonclassical effects in the entropy squeezing for the case $l=2$.

\section*{References}

\end{document}